\pdfoutput=1
\documentclass[%
 aip,jmp,amsmath,amssymb,
reprint,%
]{revtex4-1}
\usepackage{bbm}
\usepackage{graphicx}
\usepackage{dcolumn}

\usepackage{bm}
\usepackage[dvipsnames]{xcolor,colortbl}
\usepackage{mathrsfs}
\usepackage{lineno}

\begin{document}


\title{Non-reciprocal Wave Phenomena in Energy Self-reliant Gyric Metamaterials}

\author{M. A. Attarzadeh}
\author{S. Maleki}
\author{J. L. Crassidis}
\author{M. Nouh}
\email{mnouh@buffalo.edu (Corresponding author)}
\affiliation{$^1$Mechanical \& Aerospace Engineering Dept., University at Buffalo (SUNY), Buffalo, NY 14260-4400, USA}

\date{\today}

\begin{abstract}
This work presents a mechanism by which non-reciprocal wave transmission is achieved in a class of gyric metamaterial lattices with embedded rotating elements. A modulation of the device's angular momentum is obtained via prescribed rotations of a set of locally housed spinning motors and are then used to induce space-periodic, time-periodic, as well as space-time-periodic variations which influence wave propagations in distinct ways. Owing to their dependence on gyroscopic effects, such systems are able to break reciprocal wave symmetry without stiffness perturbations rendering them consistently stable as well as energy self-reliant. Dispersion patterns, band gap emergence, as well as non-reciprocal wave transmission in the space-time-periodic gyric metamaterials are predicted both analytically from the gyroscopic system dynamics as well as numerically via time-transient simulations. In addition to breaking reciprocity, we show that the energy content of a frictionless gyric metamaterial is conserved over one temporal modulation cycle enabling it to exhibit a stable response irrespective of the pumping frequency.
\end{abstract}


\maketitle

\section{Introduction}
Inertial and elastic components represent the building blocks of metamaterials and phononic crystals which have been predominantly utilized to manipulate incident waves over the past few decades \cite{brillouin2003wave, mead1996wave, deymier2013acoustic, Hussein2014}. Wave control mechanisms have thus far been limited to modulations of mass and stiffness in spatial domains \cite{huang2009negative, Nouh2015, AlBabaa2016a, Pai2010}, temporal domains \cite{cullen1958travelling}, or both \cite{trainiti2016non}. Alternatively, angular momentum arising from the rotation of distributed masses combined with the gyroscopic effect can bring about intriguing mechanical features. Examples include gyroscopic stabilization \cite{seyranian1995gyroscopic,gaudin1981gyro,barkwell1992gyroscopically}, vibration control through distributed and discrete networks of gyroscopes \cite{brocato2009control, hu2013recursive}, spacecraft attitude control \cite{yoon2002spacecraft}, shielding cloaks \cite{brun2012vortex} and, most recently, back-scattering immune edge states \cite{nash2015topological, wang2015topological}. In the latter, gyroscopic metamaterials are created by a two-dimensional network of objects rotating at a constant speed with the purpose of realizing topological edge states inspired by the Quantum Hall Effect. Contrary to dissipative elements, gyroscopic effects (herein called \textit{Gyric} after D'Eleuterio \cite{d1984dynamics}) are of a conservative nature, and do not drain the system's mechanical energy since gyroscopic forces are always orthogonal to the velocity vector. Furthermore, unlike the mass and stiffness, which are understandably difficult to modulate, angular momentum is not an inherent property of the medium but rather an outcome of a rigid body rotation (e.g.~a rotor) at a desired speed. As such, a system's angular momentum can take on different values (small or large, positive or negative) and is simply tuned by adjusting the rotation speed in real-time.

In this effort, we examine dispersion characteristics, as well as non-reciprocal wave phenomena, in a class of gyric metamaterials (GMMs) which comprise a set of embedded spinning rotors. Unlike conventional rotating periodic structures \cite{alsaffar2018band}, GMMs presented here are capable of incorporating angular momentum modulation in space, time, or both space and time independent of its constitutive mass and stiffness matrices. Gyric effects appear in the governing GMM dynamics as moments induced by a change in the angular momentum vector. A specific interest of this work is the illustration of non-reciprocal wave phenomena in GMMs stemming from breaking their wave transmission symmetry. Breaking wave reciprocity in linear systems has been a growing focus of recent acoustic and elastic metamaterials research \cite{achenbach2012wave, zanjani2014one,swinteck2015bulk}. It has been shown that the induction of linear or angular motion, whether physically \cite{fleury2014sound, attarzadeh2018elastic, beli2018mechanical} or artificially \cite{Nassar201710,attarzadeh2018wave,attarzadeh2018non} can instigate non-reciprocal wave propagation. On the realization front, some efforts have proposed the achievement of non-reciprocal systems via external stimulation of adaptive structures, e.g. using electrical fields \cite{croenne2017brillouin}, magnetic fields \cite{Ansari17}, or alternatively via torsional mechanical waves \cite{chaunsali2016stress}.

Nonetheless, most if not all of these investigations target a modulation of the system's stiffness properties, which risks influencing structural aspects. Stiffness modulations also often require hard-wiring as well as electrical shunt circuits rendering them experimentally tedious \cite{trainiti2018time}. In addition to operating with an unperturbed stiffness matrix, non-reciprocal GMMs have the ability to exhibit negative values of angular momentum $h(x,t)$ at any location $x$ or time instant $t$, simply by reversing the direction of rotation. The latter can be a cumbersome task in stiffness-modulated systems, or at least not feasible without an added layer of active control. As a result, they provide a much larger range of parameters and design flexibility. Finally, as will be shown, time-periodic GMMs reliably exhibit a stable response and a bounded energy content, irrespective of the pumping frequency required to onset the modulation; a feature which is not possessed by systems with time-periodic elastic fields \cite{cassedy1967dispersion}.

\begin{figure*}[h!]
\includegraphics[width=0.9\textwidth]{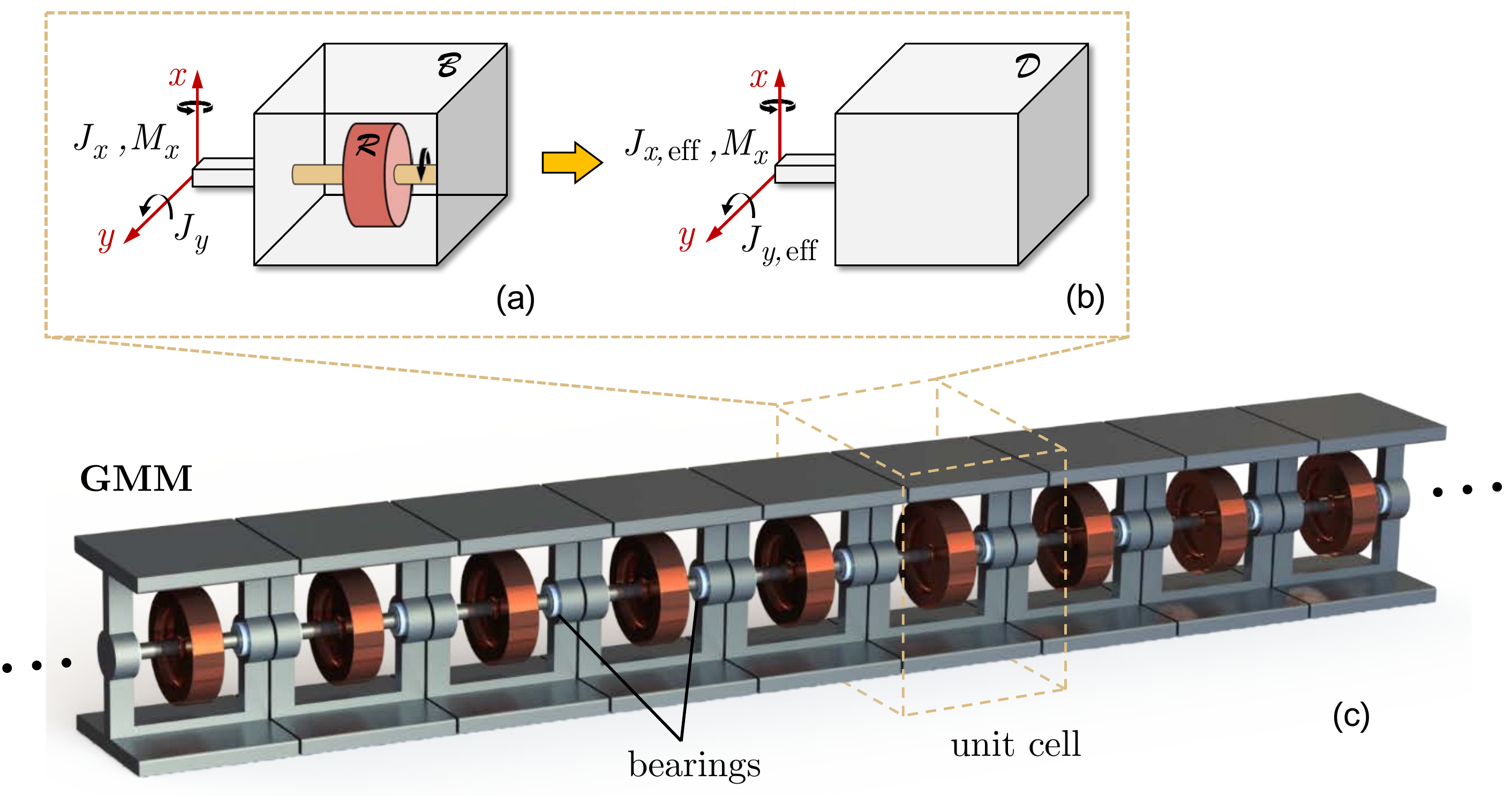}
\centering
\caption{(a) Schematic of a gyric body: Rigid body $\mathcal{B}$ with a spinning rotor $\mathcal{R}$. (b) A non-gyric body with dynamically equivalent rotational inertia. (c) GMM lattice comprising a set of interconnected gyric cells with embedded spinning rotors}
\label{fig:Gyric_Body}
\end{figure*}

\section{Structural Dynamics of a Gyric Body}
Inspired by previous efforts on gyro-elastic continua \cite{d1984dynamics} and the modified Newton's second law \cite{milton2007modifications}, we start by investigating the linear dynamics of a gyric rigid body. The body comprises a distributed mass $\mathcal{B}$ with an embedded spinning rotor $\mathcal{R}$, as depicted in Fig.~\ref{fig:Gyric_Body}a, which represents the building block of a periodic gyric structure. Unlike conventional elastic metamaterials where the outer body is connected to an internal resonator and transmits a force to it \cite{Huang2010,albabaa2017PZ}, the interaction between the outer and inner components here is rather a gyroscopic moment emerging as a result of the change in the total angular momentum vector of the gyric body (i.e. $\mathcal{B}+\mathcal{R}$). The same moment is also responsible for the precession phenomenon of a spinning top \cite{crabtree1914elementary}. The magnitude of the gyroscopic moment is proportional to $\mathcal{R}$'s angular momentum as well as $\mathcal{B}$'s rate of rotation. The direction of the moment, however, is governed by the right-hand rule and is therefore binormal to both of the rotational directions of $\mathcal{R}$ ($z$-axis) and $\mathcal{B}$ ($x$-axis). As a result, it influences the dynamics of the system in the $y$-direction. The previous instigates cross-coupling between the dynamics of the body in the $x$- and $y$-directions. We assume the system is fixed about the $z$-axis and study the two rotational degrees of freedom $\theta_x$ and $\theta_y$. In the absence of elastic (restoring) and frictional forces, the governing equations of motion are found by a simple moment balance as
\begin{subequations} \label{eq:Dyanmics_GyricBody}
\begin{align}
&J_x \ddot{\theta }_x + {h}\dot{\theta}_y=M_x\\
&J_y\ddot{\theta }_y - {h}\dot{\theta}_x=0
\end{align}
\end{subequations}

\noindent where $J_x$ and $J_y$ are the moments of inertia about the shown principal axes, and $M_x$ is the external moment applied to the outer mass in the $x$-direction. Also, $\dot{\theta}_{x,y}$ and $\ddot{\theta}_{x,y}$ represent the angular velocities and accelerations in the respective directions. The coupling term appears solely as a result of including the rotor's angular momentum $h$ in the system dynamics. The steady-state response amplitude of Eq.~(\ref{eq:Dyanmics_GyricBody}) to a harmonic moment of amplitude $\hat{M}_x$ and frequency $\omega$ is given by
\begin{equation} \label{eq:SteadyState_Amplitude_GyricBody}
    \hat{\theta}_x=\frac{\hat{M}_x}{ h^2/J_y - J_x \omega^2}
\end{equation}

\begin{figure}[h!]
\includegraphics[width=0.5\textwidth]{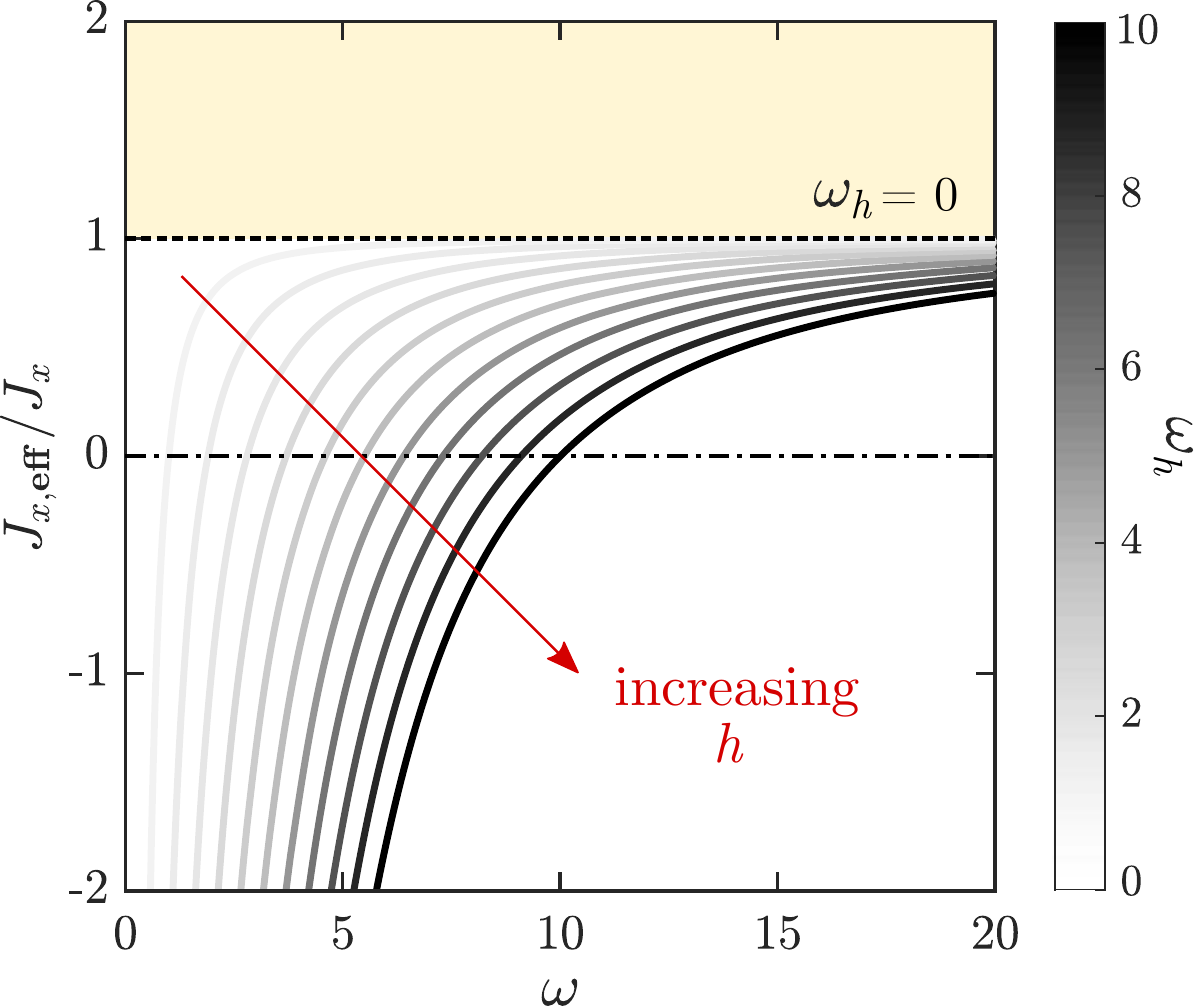}
\centering
\caption{Effective inertia $J_{x, \text{eff}}$ of a gyric body for increasing values of $\omega_h$}
\label{fig:EffectiveInertia_GyricBody}
\end{figure}

Using Eq.~(\ref{eq:SteadyState_Amplitude_GyricBody}), the gyric body can be alternatively presented by a dynamically equivalent non-gyric one ($\mathcal{D}$ in Fig.~\ref{fig:Gyric_Body}b) with an effective inertia $J_{x,\text{eff}}$, given by

\begin{equation} \label{eq:EffectiveInertia_GyricBody}
    J_{x,\text{eff}} =\bigg[1 - \frac{\omega_h^2}{\omega^2} \bigg] J_x
\end{equation}
\noindent which mimics the frequency-dependent behavior of the gyric body $\mathcal{B}$. In Eq.~(\ref{eq:EffectiveInertia_GyricBody}), $\omega_h$ denotes the equivalent rotational speed of $\mathcal{R}$ with $\omega_h^2=\frac{h^2}{J_x J_y}$. Note that a similar result could be obtained for $J_{y,\text{eff}}$ by replicating this equivalence in the $y$-direction. Fig.~\ref{fig:EffectiveInertia_GyricBody} displays variation of the effective inertia $J_{x,\text{eff}}$ with frequency for different values of $\omega_h$. The figure shows that the spinning rotor induces an artificial rotational stiffness $h^2/J_y$ in the dynamics of the system which in turn yields a negative effective inertia for frequencies in the range $0\leq\omega<\omega_h$; a phenomenon which is highly desirable in low-frequency vibroacoustic applications \cite{huang2009negative}. Given these underlying features, the rest of this work focuses on wave propagation aspects stemming from incorporating such gyric bodies in metamaterial lattices.

\section{Wave Propagation}\label{Wave Dispersion in gyric-metamaterial}
\begin{figure*}[h!]
\includegraphics[width=0.9\textwidth]{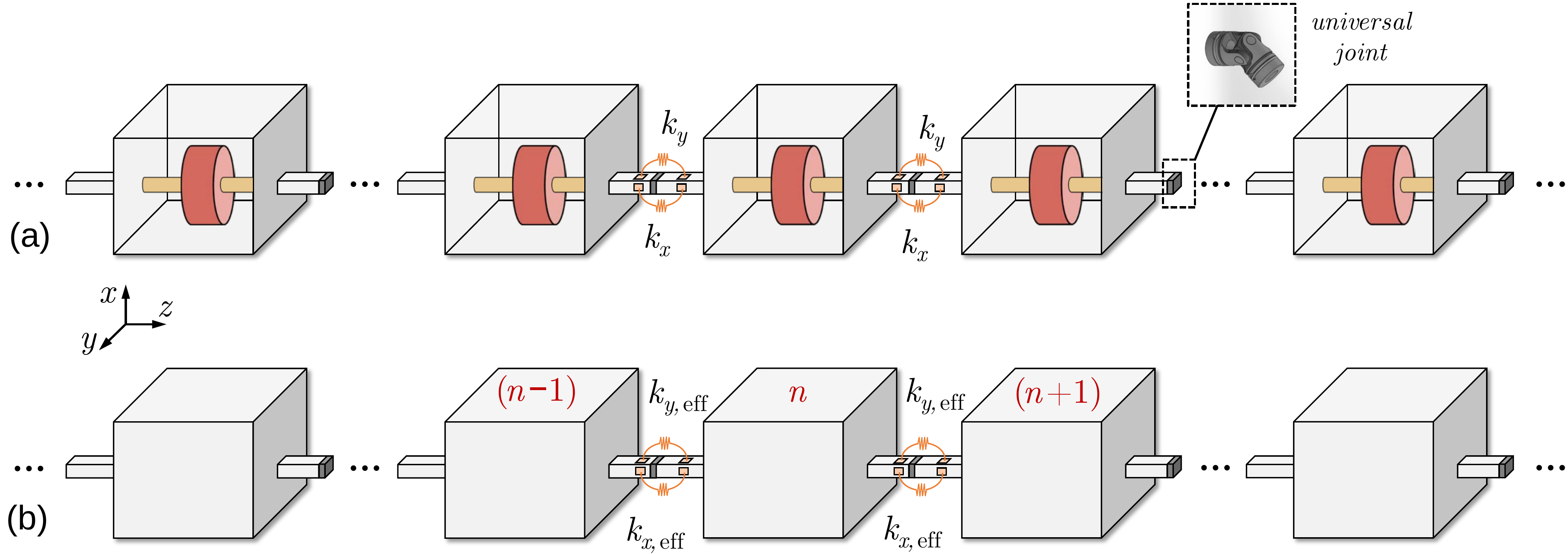}
\centering
\caption{(a) Schematic of a GMM lattice. 2-DOF universal joints and torsional springs across the $x-$ and $y-$ axes connect the adjacent unit cells. (b) A dynamically equivalent non-gyric lattice with a frequency-dependent effective stiffness}
\label{fig:Lattice}
\end{figure*}

In this section, different types of gyric lattices are devised and studied in order to explore the full potential of GMMs in manipulating incident elastic waves. In its most general case, a GMM is formed by connecting multiple gyric unit cells with torsional springs, such as shown in Fig.~\ref{fig:Lattice}a. In the provided schematic, every unit cell is connected to its adjacent ones through a universal joint allowing rotations in the two lateral directions $x$ and $y$ only and is supported with torsional springs in both directions. The entire lattice is fixed about the local $z$-axis. Throughout this analysis, the fundamental mechanical properties, i.e. stiffness and inertia, are kept completely unchanged in the different proposed GMM configurations, in addition to being invariant spatially and temporally. Instead, the angular momentum $h$ is varied in different ways to onset band gaps, as well as break wave reciprocity and transmission symmetry. As a result, such GMMs provide the additional advantage of real-time tunability as well as a switchable platform between different functionalities, as will be detailed. 

Although asymmetric elastic and inertial properties about the $x$- and $y$-axes might provide an additional degree of design flexibility, it is rather challenging to control these properties in real-time. Therefore, we limit our analysis to $k_x=k_y=k$ and $J_x=J_y=J$ in order to isolate the angular momentum role in tuning the dispersion profile from other parameters. In the following subsections, we discuss the dispersion of elastic waves in GMMs with uniform, space-periodic, time-periodic and space-time-periodic angular momentum distributions, and follow that with a numerical time-transient verification of the fully simulated system in the last section.

\subsection{Uniform Angular Momentum Distribution \label{Uniform Angular Momentum Distribution}} 
Consider an infinite lattice made of gyric unit cells, as shown in Fig.~\ref{fig:Lattice}a, where the index $n$ specifies the global position of each cell on the $z$-axis with respect to the origin. We start by setting an identical and time-invariant $h$ for all cells. Consequently, the $n^{\text{th}}$ unit cell equations in the frequency domain can be written as
\begin{subequations} \label{eq:Dyanmics_Uniform}
\begin{align}
- J \omega^2 {\theta}_x^{(n)} + i \omega h {\theta}_y^{(n)} + k   \big(2\theta_x^{(n)} -\theta_x^{(n-1)}-\theta_x^{(n+1)} \big)=0\\
-J \omega^2 {\theta}_y^{(n)} - i \omega h {\theta}_x^{(n)} + k\big(2\theta _y^{(n)} -\theta _y^{(n-1)}-\theta _y^{(n+1)}\big)=0
\end{align}
\end{subequations}
\noindent where $i=\sqrt{-1}$.  Owing to the periodicity of lattice, a solution of the form $\theta_{x,y}^{(n)}=\hat{\theta}_{x,y} e^{- i n \kappa}$ can be adopted, where $\kappa$ is the non-dimensional wave number. Substituting the solution back in Eq.~(\ref{eq:Dyanmics_Uniform}) gives
\begin{equation} \label{eq:DispersionRelation_Uniform}
    \cos\kappa=1- \frac{1}{2}(\Omega^2 \pm  \Omega_h \Omega)   
\end{equation}
\noindent with the corresponding Bloch modes (eigenvectors):
\begin{equation} \label{eq:Eigenvectors_Uniform}
    \begin{bmatrix}
    \hat{\theta}_x \\ \hat{\theta}_y\\ \end{bmatrix} =\begin{bmatrix}
    1 \\ \pm i
    \end{bmatrix}   
\end{equation}
\noindent where $\Omega=\frac{\omega}{\omega_0}$ is the dimensionless frequency, $\Omega_h=\frac{\omega_h}{\omega_0}$ is the dimensionless angular momentum of the spinning rotors, and $\omega_0=\sqrt{k/J}$ is the natural frequency. The group velocity at the long wavelength limit ($c_g=d\Omega/d \kappa$) is zero, indicating that the introduction of angular momentum results in a vanishing speed of sound in GMMs with a uniform $h$ distribution. We also note that the $\pm$ appears in both Eqs.~(\ref{eq:DispersionRelation_Uniform}) and (\ref{eq:Eigenvectors_Uniform}) because two distinct wavenumbers ($\kappa_{+}$ and $\kappa_{-}$) can coexist in each Irreducible Brillouin Zone (IBZ) for every incident frequency $\Omega$. The Bloch modes in Eq.~(\ref{eq:Eigenvectors_Uniform}) reveal that transverse deflections about the $x$- and $y$- axes have a $\pm \frac{\pi}{2}$ phase shift corresponding to each of the $[1,-i]$ and $[1, i]$ modes, respectively. In essence, for each mode, two coupled orthogonal transverse waves travel in the spanwise direction of a 1D medium with identical wavenumbers and magnitudes. Henceforth, we refer to each of these coupled transverse wave pairs as a \textit{wave-mode}. Therefore, the two wavenumbers $\kappa_{+}$ and $\kappa_{-}$ correspond to two separate wave-modes propagating in the GMM, both expected to materialize in the lattice simultaneously for any given excitation.

\begin{figure*}[h!]
\includegraphics[width=\textwidth]{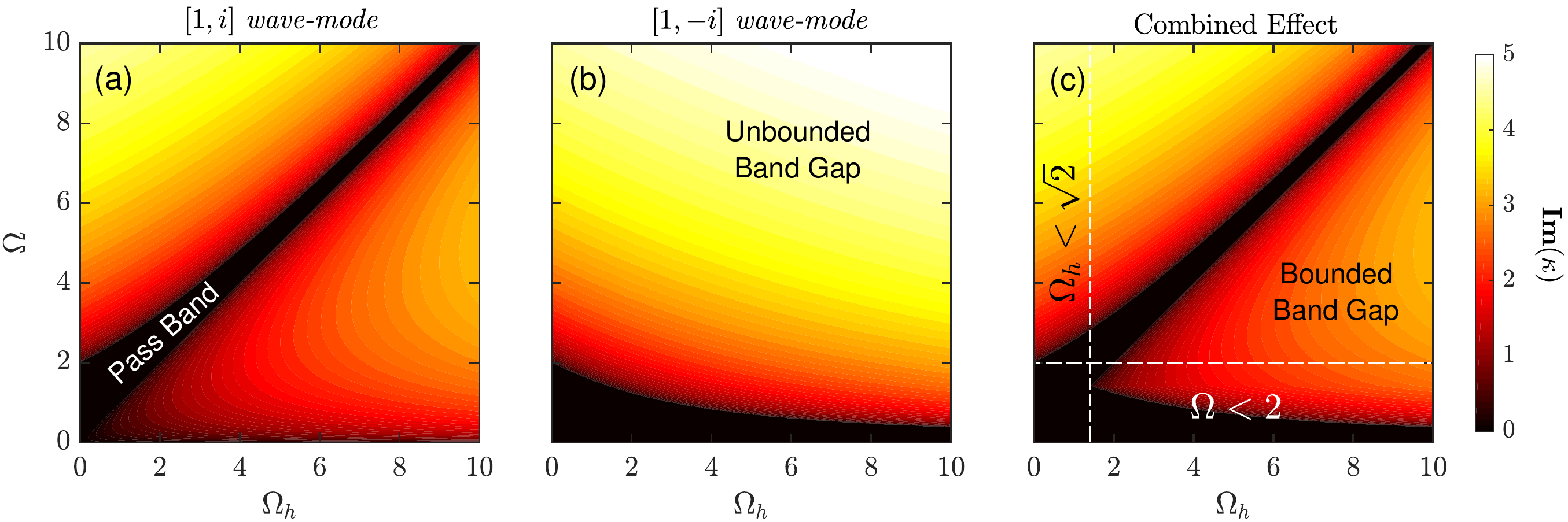}
\centering
\caption{Attenuation degree $\mathbf{Im}(\kappa)$ as a function of $\Omega_h$ corresponding to: (a) The $[1,i]$ wave-mode, (b) the $[1,-i]$ wave-mode, and (c) the combined effect showing complete band gaps. A bounded band gap exists only for $\Omega_h >\sqrt{2}$}
\label{fig:Attenuation_Uniform}
\end{figure*}

A frequency band gap forms in the range where $\kappa$ is complex-valued. As such, given Eq.~(\ref{eq:DispersionRelation_Uniform}), it is evident that the $[1, i]$ mode is attenuated in the ranges $0 < \Omega < \Omega_h$ and $\Omega > \frac{1}{2}(\sqrt{16+\Omega_h^2}+\Omega_h)$, while the $[1,-i]$ mode is attenuated only if $\Omega > \frac{1}{2}(\sqrt{16+\Omega_h^2}-\Omega_h)$. Each of these cases corresponds to a partial band gap where one mode attenuates and the other freely propagates. A complete band gap, where both modes are suppressed, occurs in the shared frequency range $\frac{1}{2}(\sqrt{16+\Omega_h^2}-\Omega_h) < \Omega < \Omega_h$ and $\Omega > \frac{1}{2}(\sqrt{16+\Omega_h^2}+\Omega_h)$. These ranges as well as the corresponding attenuation degree, i.e. $\textbf{Im}(\kappa)$, are graphically illustrated in Fig.~\ref{fig:Attenuation_Uniform} as a function of the spinning rotor angular momentum $\Omega_h$. As can be seen in Fig.~\ref{fig:Attenuation_Uniform}c, the GMM maintains a bounded band gap only if $\Omega_h<\sqrt{2}$. Further, for the limiting case of $\Omega_h=0$, the unbounded band gap (commonly referred to as a \textit{stop band} \cite{Hussein2014}) occurs at $\Omega > 2$ which, as anticipated, matches the stop band of a discrete monatomic dispersive structure \cite{brillouin2003wave}. 

\begin{figure*}
\includegraphics[width=\textwidth]{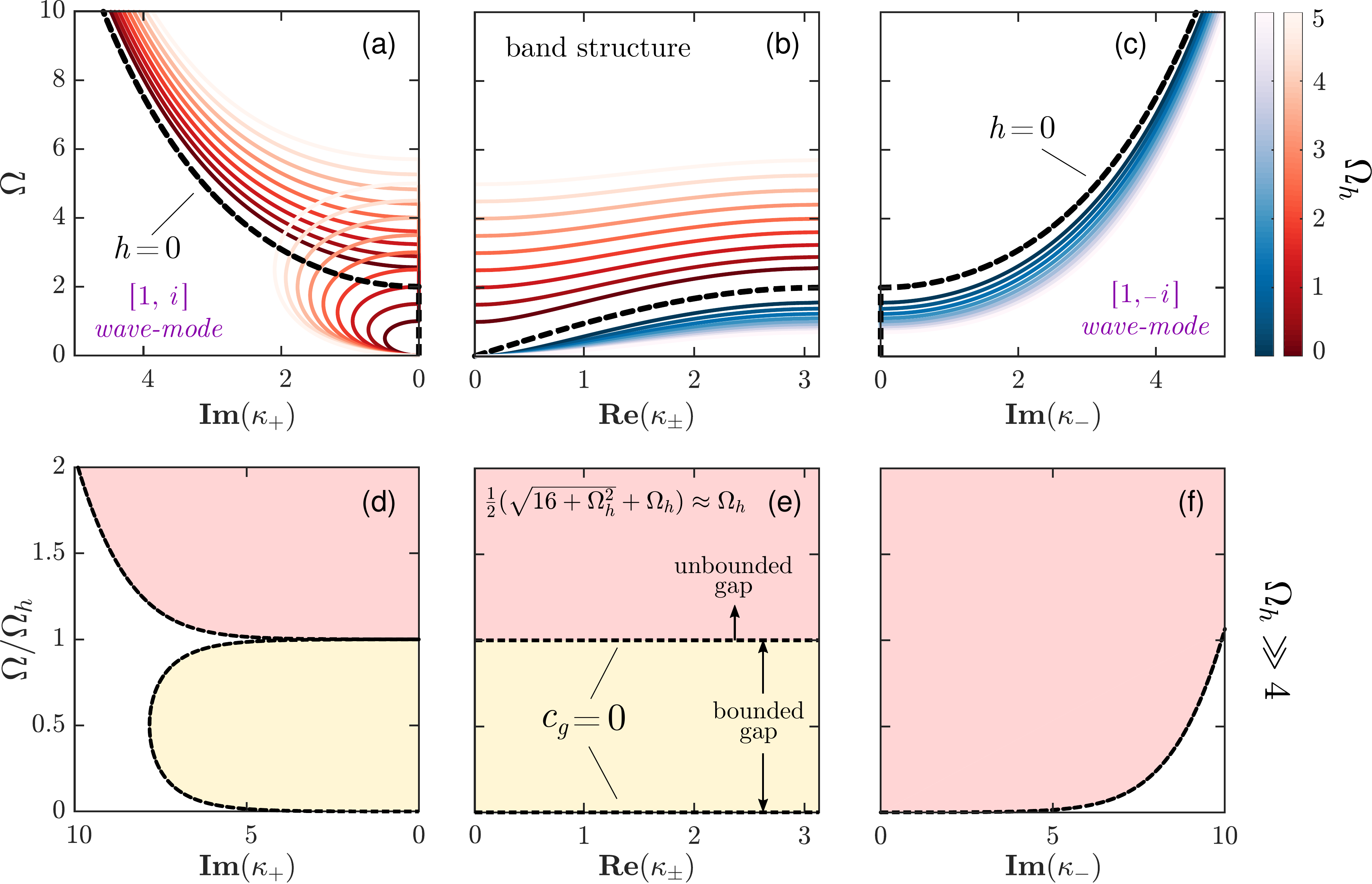}
\centering
\caption{(a-c) Dispersion diagrams for a GMM with a uniform angular momentum in the frequency range $0<\Omega_h<5$. Color levels indicate the value of $\Omega_h$. Dashed-line corresponds to a non-gyric monatomic lattice ($h=0$). (d-f) Dispersion diagrams for $\Omega_h = 50$ showing the localized wave modes at $\Omega=0$ and $\Omega=\Omega_h$ for the case when $\Omega_h \gg 4$}
\label{fig:Dispersion_Uniform}
\end{figure*}

The previous characteristics are also confirmed by the GMM's dispersion diagrams shown in Fig.~\ref{fig:Dispersion_Uniform}a. The middle panel represents the unit cell's band structure given by the real component of the wavenumbers while the leftmost and rightmost panels show the corresponding attenuation degrees $\mathbf{Im}(\kappa_+)$ and $\mathbf{Im}(\kappa_-)$, respectively. The dashed lines correspond to zero angular momentum ($h=0$) or a non-gyric structure where $\mathbf{Im}(\kappa_+)=\mathbf{Im}(\kappa_-)$. More importantly, we note that as $\Omega_h$ increases, the bounded band gap forms in the leftmost panel which corresponds to the $[1, i]$ mode. This signals that the two degenerate modes of the system start diverging until they eventually become flat single frequency bands at $\Omega=0$ and $\Omega=\Omega_h$, as depicted in Fig.~\ref{fig:Dispersion_Uniform}e. In which case, the bounded and unbounded band gaps coalesce to span the entire dispersion spectrum with the exception of these two frequencies. Upon examining the band gap ranges discussed earlier, we expect this limiting behavior to happen for $\Omega_h \gg 4$ where $\frac{1}{2}(\sqrt{16+\Omega_h^2}-\Omega_h) \approx 0$ and  $\frac{1}{2}(\sqrt{16+\Omega_h^2}+\Omega_h) \approx \Omega_h$. In practice, such localized flat bands, with near-zero group velocities, behave similar to standing waves associated with natural frequencies of a finite structure. As a result, the GMM manages to selectively admit both frequencies from a wide-band excitation. Such behavior has been also recently reported in time-periodic systems where \textit{wavenumber} as opposed to \textit{frequency} band gaps emerge as a result of an external temporal modulation \cite{trainiti2018time}. It is also noteworthy that, given the dependence on $h$ in GMMs, this apparent band separation can be tuned in real-time by solely varying the rotational speed of the rotors. Finally, by reversing the rotation direction of $\mathcal{R}$, the dispersion profile switches between the wave-modes. Substituting $\Omega_h$ with $-\Omega_h$ does not influence the band structure shown in Fig.~\ref{fig:Dispersion_Uniform} (with the exception of swapping the subplots of $\textbf{Im}(\kappa_+)$ and $\textbf{Im}(\kappa_-)$) while perfectly exchanging the two eigenvectors. The latter being an added feature of gyroscopic systems, which has been recently utilized to induce internal cell asymmetry between neighboring unit cells thus creating topologically protected interface modes based on the quantum valley hall effect (QVHE) \cite{garau2018interfacial}.

To shed light on the band gap generation mechanism in GMMs, consider a lattice made of the equivalent non-gyric unit cells as shown in Fig.~\ref{fig:Lattice}b. To establish dynamic equivalence, the non-gyric lattice exhibits a frequency-dependent effective stiffness $k_{\text{eff}}$ to compensate for the embedded spinning rotor in its gyric counterpart. The dispersion relation for such lattice becomes
\begin{equation} \label{eq:DispersionRelation_EquivalentLattice}
    \cos\kappa =1- \frac{J \omega^2}{2k_{\text{eff}}}
\end{equation}
\noindent and by comparing Eqs.~(\ref{eq:DispersionRelation_Uniform}) and (\ref{eq:DispersionRelation_EquivalentLattice}) and solving for $k_{\text{eff}}$, we get
\begin{equation} \label{eq:EffectiveStiffness_Uniform}
k_{\text{eff}}=\frac{k}{1 \pm \frac{\Omega_h}{\Omega}} 
\end{equation}

\begin{figure}[h!]
\includegraphics[width=0.5\textwidth]{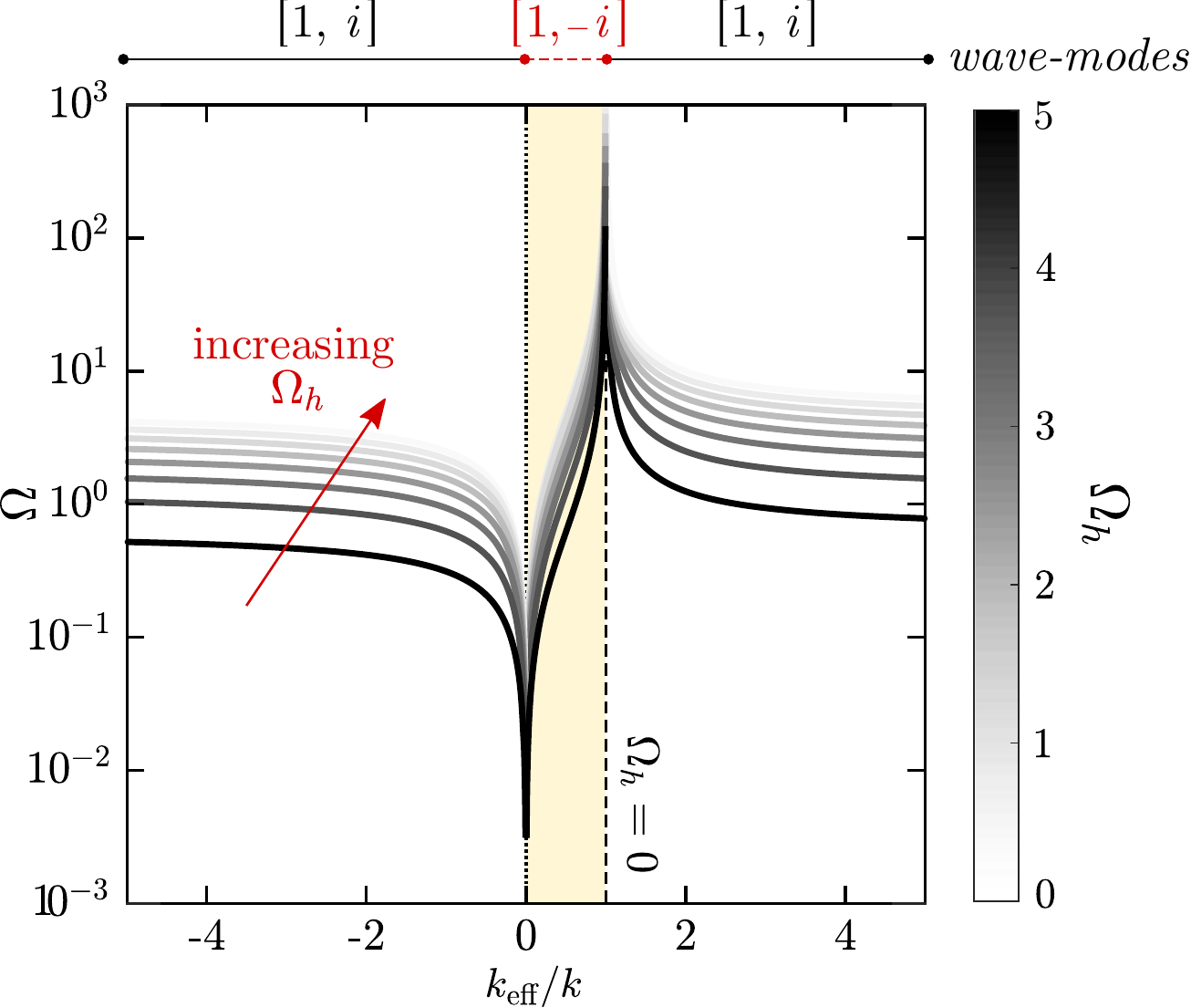}
\centering
\caption{Effective stiffness properties of a dynamically equivalent non-gyric lattice for increasing values of $\Omega_h$}
\label{fig:EffectiveStiffness_Uniform}
\end{figure}

Fig.~\ref{fig:EffectiveStiffness_Uniform} captures the relationship between $\Omega$ and $k_{\text{eff}}$ for different values of $\Omega_h$. It confirms that the stiffness of the equivalent non-gyric lattice exhibits negative effective values within the range $0<\Omega<\Omega_h$ for the $[1, i]$ mode only. Further, the negativity switches between the the two wave-modes with an angular momentum sign change, which correlates well with the predicted band gaps.

\subsection{Space-Periodic Angular Momentum Variation \label{space-time}}
Next, we examine a phononic GMM with a spatially periodic pattern by setting the rotor speed of every other cell equal to zero. In which case, we are able to model the steady-state dynamics of the $n^{\text{th}}$ and $(n-1)^{\text{th}}$ successive cells via the system of equations given by
\begin{equation}
\mathbf{K}_d \hspace{0.1cm} \bm{\Theta} - k \bm{\Theta_r} = \mathbf{0}
\label{eq:Dyanmics_Space_Periodic}
\end{equation}
\noindent where
\begin{equation}
    \mathbf{K}_d =
    \begin{bmatrix}
    2k-\omega^2 J & 0 & -k & 0\\
    0 & 2k-\omega^2 J & 0 & -k \\
    -k & 0 & 2k-\omega^2 J & i \omega h\\
    0 & -k & - i \omega h & 2k-\omega^2 J
    \end{bmatrix}
\end{equation}
\begin{equation}
    \bm{\Theta} =
    \begin{bmatrix}
    \theta_x^{(n-1)} & \theta_y^{(n-1)} & \theta_x^{(n)} & \theta_y^{(n)}
    \end{bmatrix} ^\mathrm{T}
\end{equation}
\noindent and
\begin{equation}
    \bm{\Theta_r} =
    \begin{bmatrix}
    \theta_x^{(n-2)} & \theta_y^{(n-2)} & \theta_x^{(n+1)} & \theta_y^{(n+1)}
    \end{bmatrix} ^\mathrm{T}
    \label{eq:theta_r}
\end{equation}

Eqs.~(\ref{eq:Dyanmics_Space_Periodic}) through (\ref{eq:theta_r}) constitute four coupled mechanical resonators. As a result, we expect at most four different modes to appear in the system's response. Similar to the previous section, the number of modes is reduced by half for a vanishing angular momentum (i.e. $h=0$). Given the GMM's periodicity, a Bloch-wave solution of the form $\theta_{x,y}^{(n-2)}=\theta_{x,y}^{(n)} e^{i \kappa}$ and $\theta_{x,y}^{(n+1)}=\theta_{x,y}^{(n-1)} e^{-i \kappa}$ is employed. Upon substituting in Eq.~(\ref{eq:Dyanmics_Space_Periodic}), we obtain the dispersion relation
\begin{equation} \label{eq:Dispersion_Relation_Space_Periodic}
    \cos \kappa =\frac{1}{2} (\Omega^2-2)(\Omega^2 \pm \Omega_h \Omega -2) -1
\end{equation}

\begin{figure*}[h!]
\includegraphics[width=\textwidth]{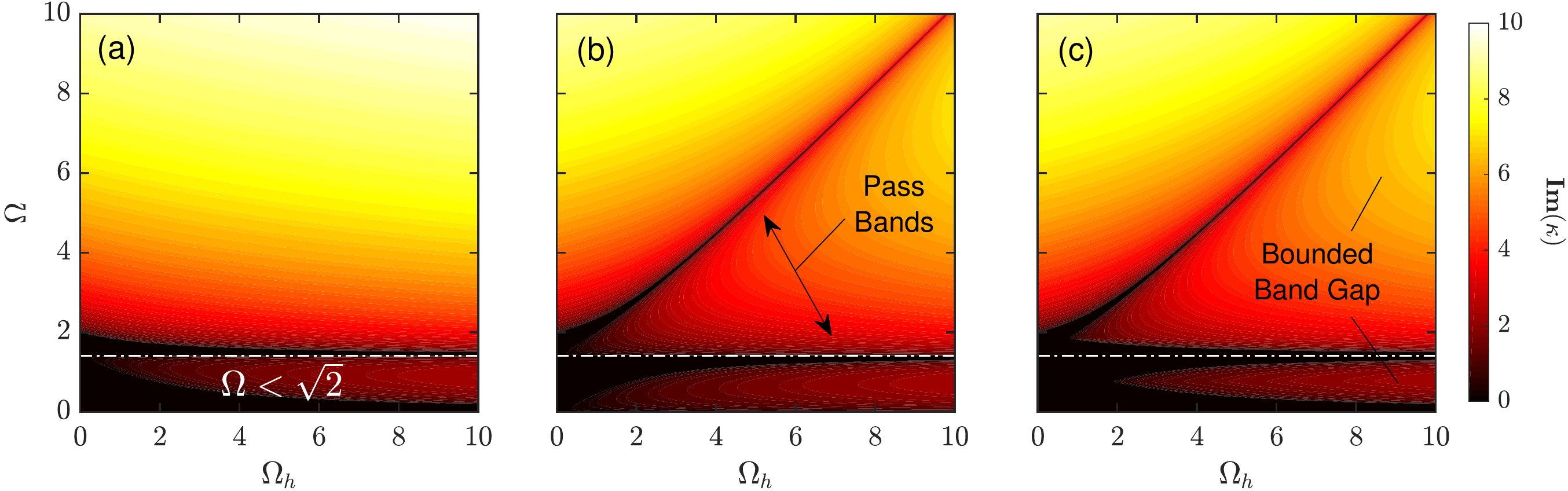}
\centering
\caption{(a-b) Attenuation degree $\mathbf{Im}(\kappa)$ as a function of $\Omega_h$ for partial band gaps. (c) Shared attenuation regions showing three complete band gaps (two of which are bounded) for a space-periodic GMM}
\label{fig:Attenuation_SpacePeriodic}
\end{figure*}

\begin{figure*}[h!]
\includegraphics[width=\textwidth]{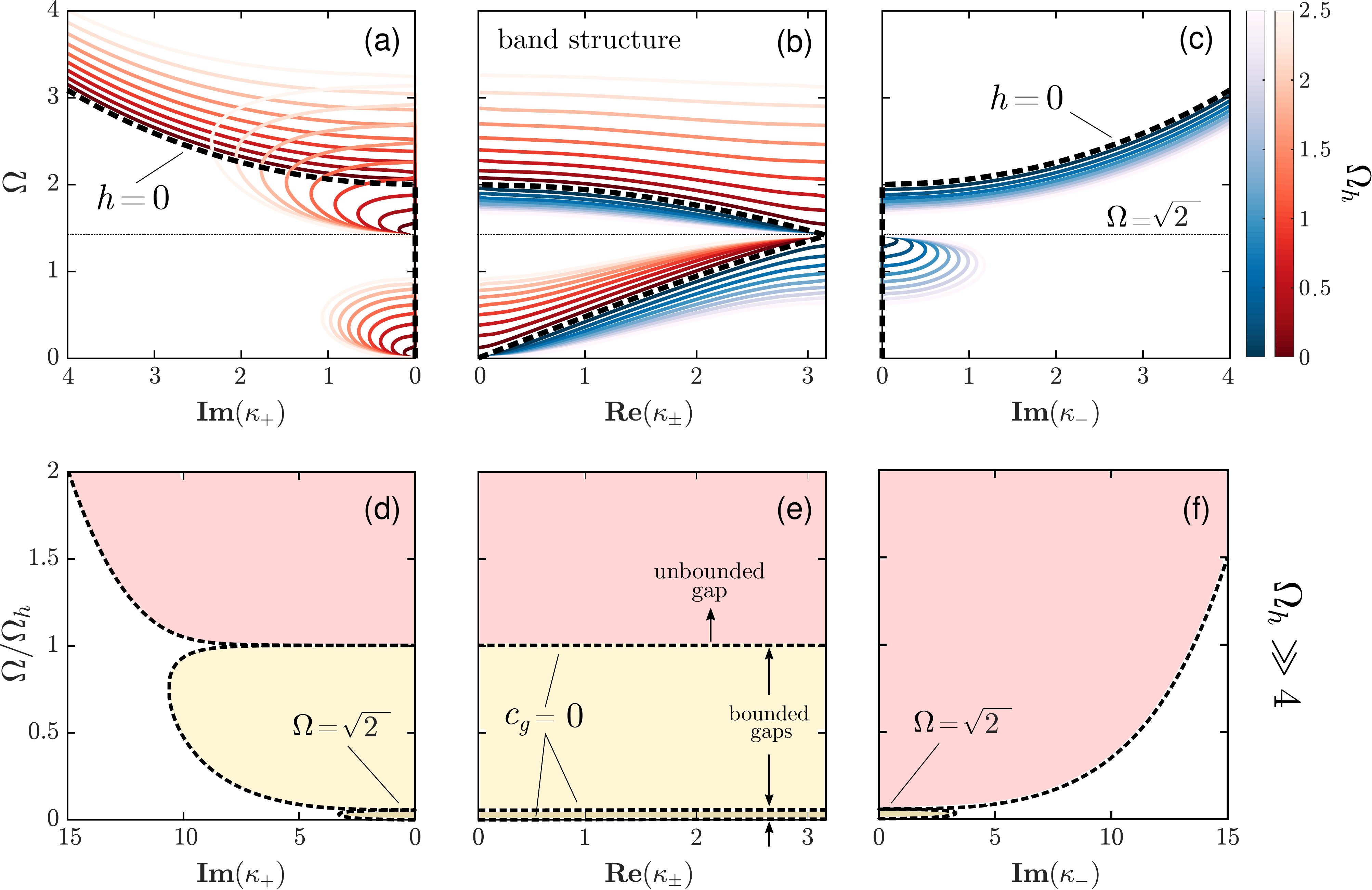}
\centering
\caption{(a-c) Dispersion diagrams for a GMM with a space-periodic angular momentum distribution within the frequency range $0<\Omega_h<2.5$. Color levels indicate the value of $\Omega_h$. Dashed-line corresponds to a non-gyric lattice ($h=0$). (d-f) Dispersion diagrams for $\Omega_h=25$ showing the localized wave modes at $\Omega=0$, $\Omega=\sqrt{2}$, and $\Omega=\Omega_h$ for the case when $\Omega_h \gg 4$}
\label{fig:Dispersion_SpacePeriodic}
\end{figure*}

It is evident that the $\pm \Omega_h \Omega$ term in Eq.~(\ref{eq:Dispersion_Relation_Space_Periodic}) vanishes as $\Omega_h$ approaches zero, causing a reduction in distinct wave modes. Band gap regions and attenuation degrees for different values of $\Omega_h$ are shown in Fig.~\ref{fig:Attenuation_SpacePeriodic}, with the corresponding dispersion diagrams in Fig.~\ref{fig:Dispersion_SpacePeriodic}. As predicted, the space-periodic GMM exhibits four distinct wave-modes resulting from the split of each of the acoustic and optic modes. The dashed lines correspond to a zero angular momentum value associated with a conventional non-gyric lattice with acoustic and optic modes only and no band gaps. The bottom panel of Fig.~\ref{fig:Dispersion_SpacePeriodic} depicts the large angular momentum scenario where the GMM effectively acts as a wide-band filter for all frequencies with the exception of $\Omega=0$, $\Omega=\sqrt{2}$ and $\Omega=\Omega_h$ (represented by the flat branches in the figure). In other words, the lower band approaches zero, the upper one tends  to $\Omega=\Omega_h$, and the two inner modes converge to a single band at $\Omega=\sqrt{2}$.

\subsection{Time-Periodic Angular Momentum Variation}
Inspired by efforts in utilizing time modulated material properties for frequency conversion and amplification \cite{cullen1958travelling, tien1958parametric, trainiti2018time}, we next examine a time-harmonic GMM obtained by imposing an oscillating $h$ with a pumping frequency $\omega_p$ across the entire gyric lattice. The time variation is given by $h(t)=h_0 + h_1 \cos \omega_p t$, where $h_0$ and $h_1$ are the average (bias) and oscillation amplitudes, respectively. Consider a time-periodic GMM realized by means of a mechanical torque imparted to each spinning rotor and given by $T(t)=d h/dt=-h_1 \omega_p \sin \omega_p t$. The torque can, for example, be supplied by an internally embedded motor with an instantaneous power
\begin{equation}
\dot{P}_{\text{e}}(t)=\frac{h T}{J_r}=\frac{-\omega_p h_1}{2J_r}(2h_0 \sin\omega_p t+ h_1 \sin 2\omega_p t)
\end{equation}
\noindent where $J_r$ is the rotational inertia of the rotor about its principle $z$-axis. As such, the required power to overcome the rotational inertia of the rotor is approximately upper-bounded by $\dot{P}_{\text{e}} \cong \omega_p h_1(2h_0+ h_1)/{2J_r}$ (an exact upper bound exists, but is omitted here for brevity). For a frictionless setup, the net amount of energy injected into the system per one temporal cycle is
\begin{equation} \label{eq:Energy_TimePeriodic}
P_{\text{e/Cycle}}=\int_0^{\frac{2 \pi}{\omega_p}} \dot{P}_{\text{e}} dt =0
\end{equation}
\noindent independent of $\omega_p$, which can be attributed to the $\frac{\pi}{2}$ phase shift between the angular momentum and the electromotor torque. This behavior is reminiscent of forced harmonic oscillations of an undamped spring-mass system, where zero energy input is required to maintain oscillations at a desired frequency. It should be noted, however, that in the presence of friction or damping, a net positive amount of energy is needed to sustain the time-periodic angular momentum profile.

\begin{figure*}
\includegraphics[width=0.9\textwidth]{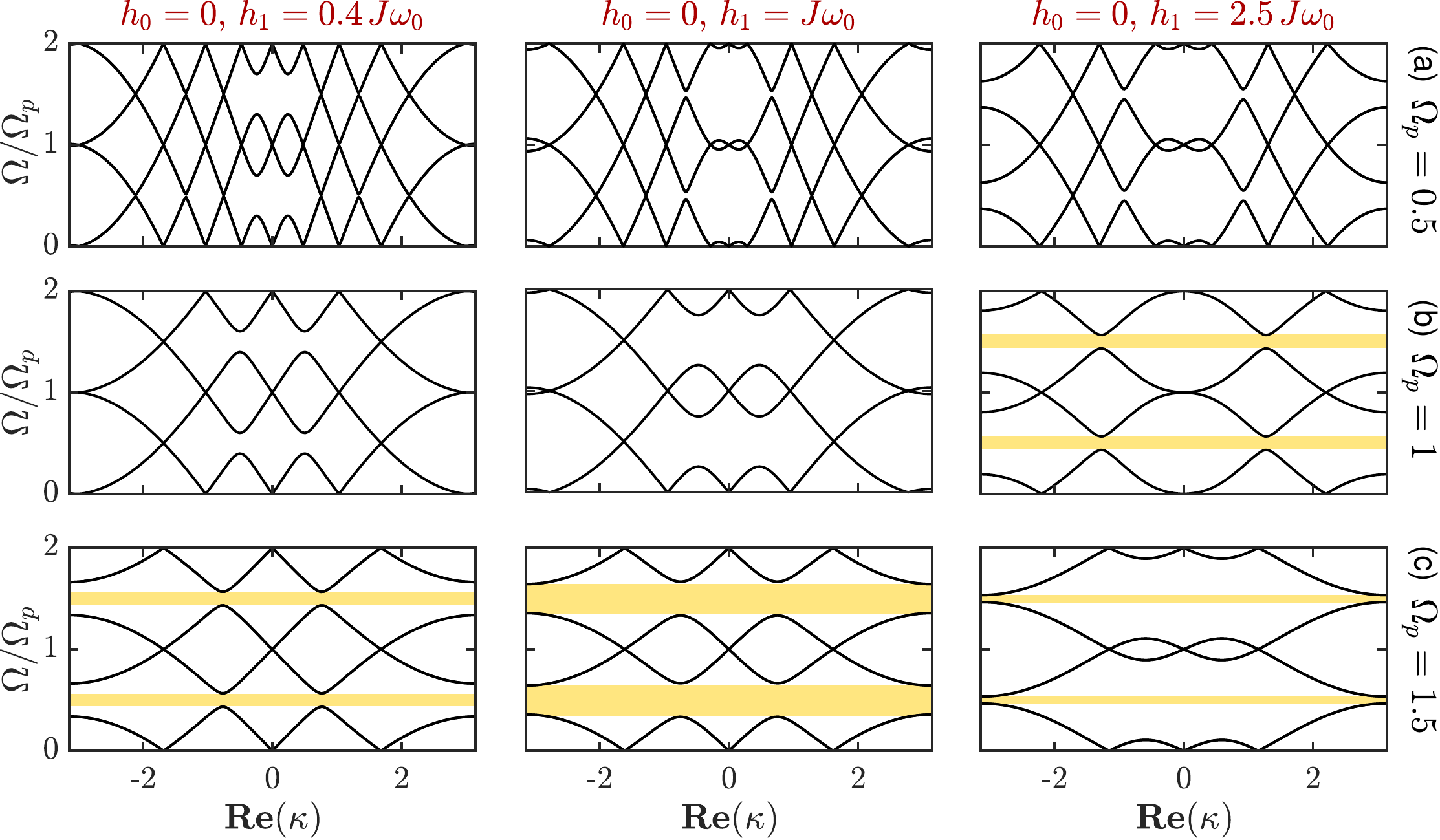}
\centering
\caption{Dispersion diagrams of a time-periodic GMM with a zero modulation bias ($h_0=0$): (a)  $\Omega_p=0.5$, (b) $\Omega_p=1.0$, and (c) $\Omega_p=1.5$. First column: $h_1=0.4 J \omega_0$, second column: $h_1=J \omega_0$, and third column: $h_1=2.5 J \omega_0$. Shaded regions indicate band gap emergence at increased pumping frequencies}
\label{fig:Dispersion1_TimePeriodic}
\end{figure*}

The time-periodic GMM's motion equations are given by
\begin{subequations} \label{eq:Dyanmics_Time_Periodic}
\begin{align}
J \ddot{\theta }_x^{(n)}= -h(t)\dot{\theta} _y^{(n)} - k \big[2\theta_x^{(n)}-\theta_x^{(n-1)}-\theta_x^{(n+1)} &\big]\\
J \ddot{\theta }_y^{(n)}= h(t)\dot{\theta} _x^{(n)} - k \big[2\theta_y^{(n)}-\theta_y^{(n-1)}-\theta_y^{(n+1)} &\big]
\end{align}
\end{subequations}

In deriving Eq.~(\ref{eq:Dyanmics_Time_Periodic}), we still maintain uniform stiffness and inertia along both transverse axes and the spanwise direction of the GMM lattice. We also assume that the inertia of the unit cells are not significantly affected by the temporal variation of the rotor's angular momentum. In practice, this is achieved by tuning the angular momentum via controlling the speed, rather than inertia, of the spinning rotors. By taking spatial periodicity into account, i.e. $\theta_{x,y}^{(n+1)} (t)=e^{-i\kappa} \theta_{x,y}^{(n)} (t)$, and $ \theta_{x,y}^{(n-1)} (t) =e^{i\kappa} \theta_{x,y}^{(n)} (t)$, and dropping $n$, Eq.~(\ref{eq:Dyanmics_Time_Periodic}) simplifies to
\begin{subequations} \label{eq:Dyanmics_Time_Periodic2}
\begin{align}
J \ddot{\theta }_x+h(t)\dot{\theta} _y + 2k (1-\cos \kappa) {\theta }_x=0\\
J \ddot{\theta }_y - h(t)\dot{\theta}_x + 2k (1-\cos \kappa) {\theta }_x= 0
\end{align}
\end{subequations}

Eq.~(\ref{eq:Dyanmics_Time_Periodic2}) constitutes a homogeneous linear time-periodic (LTP) system with a period $\tau_p=\frac{2\pi}{\omega_p}$. For any given $\kappa$, a proper ansatz given by the Floquet theory is \cite{wereley1990analysis}
\begin{equation} \label{eq:Ansatz_Time_Periodic}
     \theta_{x,y} (t;\kappa) = e^{\lambda t} \sum_{l=-\infty}^{\infty} \phi_{x,y}^{(l)} e ^{i l \omega_p t}
\end{equation}
\noindent where $\phi_{x}^{(l)}$ and $\phi_{y}^{(l)}$ are the Fourier coefficients and $\lambda$ is a generally complex constant known as the characteristic exponent \cite{richards2012analysis}. Substituting Eq.~(\ref{eq:Ansatz_Time_Periodic}) in (\ref{eq:Dyanmics_Time_Periodic2}) and exploiting harmonic balance results into a second order eigenvalue problem in terms of $\lambda$ and $\kappa$. According to the Lyapanov-Floquet theorem \cite{brockett2015finite}, the temporal nature --and thus stability \cite{xie2006dynamic}-- of the wave propagation is dictated by the characteristic exponents. As a result, it is beneficial to consider the free-wave case by casting the dispersion relation as a quadratic eigenvalue problem in terms of $\Lambda=\lambda/\omega_0$
\begin{equation} \label{eq:DispersionRelation_TimePeriodic}
\begin{Bmatrix}
\begin{bmatrix}
\mathbf{I} & \mathbf{O} \\
\mathbf{O} & \mathbf{I} \\
\end{bmatrix} \Lambda^2 +  \begin{bmatrix}
\mathbf{\Psi}^{(1)} &  \mathbf{\Psi}^{(2)}\\
-\mathbf{\Psi}^{(2)} & \mathbf{\Psi}^{(1)} \\
\end{bmatrix} \Lambda + \begin{bmatrix}
\mathbf{\Psi}^{(3)}   & \mathbf{\Psi}^{(4)} \\
- \mathbf{\Psi}^{(4)} & \mathbf{\Psi}^{(3)} \\
\end{bmatrix}
\end{Bmatrix}
\begin{Bmatrix}
\mathbf{\Phi_x} \\
\mathbf{\Phi_y} \\
\end{Bmatrix} = 0
\end{equation}
\noindent where $\mathbf{I}$ and $\mathbf{O}$ are identity and null matrices of a proper size. The entries of the dimensionless matrices $\mathbf{\Psi}^{(j)}$ for $j=1,\dots,4$ are given by
\begin{subequations} \label{eq:PSI_Time_Periodic}
\begin{align}
&\mathbf{\Psi}^{(1)}_{l,q}=2 i q \Omega_p \delta_{l,q}\\
&\mathbf{\Psi}^{(2)}_{l,q}=\frac{h_0}{J \omega_0} \delta_{l,q}+\frac{h_1}{2J \omega_0} \delta_{l,q\pm1}\\
&\mathbf{\Psi}^{(3)}_{l,q}= \big [ 2(1-\cos \kappa) - q^2 \Omega_p^2  \big ] \delta_{l,q}\\
&\mathbf{\Psi}^{(4)}_{l,q}= i q \Omega_p \mathbf{\Psi}^{(2)}_{l,q}
\end{align}
\end{subequations}
\noindent where $\Omega_p=\omega_p/\omega_0$ is the dimensionless temporal modulation frequency, $l$ and $q \in \mathbb{Z}$, and $\delta_{l,q}$ is the Kronecker delta which is equal to one for $l = q$ and zero otherwise. In addition, the rotation vectors ($\mathbf{\Phi}_x$ and $\mathbf{\Phi}_y$) are
\begin{equation} \label{eq:RotationVectors_TimePeriodic}
\mathbf{\Phi}_{x,y}=\Big[\phi_{x,y}^{(-\infty)},\dots ,\phi_{x,y}^{(-1)},\phi_{x,y}^{(0)},\phi_{x,y}^{(1)}, \dots , \phi_{x,y}^{(\infty)}\Big]^{\mathrm{T}}
\end{equation}

The vector in Eq.~(\ref{eq:RotationVectors_TimePeriodic}) can be truncated as long as the summation in Eq.~(\ref{eq:Ansatz_Time_Periodic}) remains bounded; a guaranteed property of a Fourier series provided a sonic blow-up does not happen \cite{cassedy1963dispersion}. In other words, approximate dispersion diagrams can be generated by seeking non-trivial solutions of the truncated version of Eq.~(\ref{eq:DispersionRelation_TimePeriodic}). Note that the appearance of $\mathbf{\Psi}^{(1)}$ and $\mathbf{\Psi}^{(4)}$ is solely attributed to the time-periodic angular momentum variation. The stiffness matrix of the system is also influenced since a centripetal term is subtracted from the diagonal terms in the third matrix, as evident by Eq.~(\ref{eq:PSI_Time_Periodic}c). By setting $\Omega_p =0$, we should reconstruct the problem outlined in section \ref{Uniform Angular Momentum Distribution}. 

As previously stated, unlike stiffness and density in non-gyric metastructures, the angular momentum $h$ is not limited to positive values. Consequently, time-periodic GMMs can be generally classified into three groups based on the modulation amplitudes $h_0$ and $h_1$. The first group comprises GMMs with zero bias angular momentum ($h_0=0$) and $h_1>0$. The second group is systems with a nonzero bias ($h_0 > 0$) and $h_1<h_0$. The third group has $h_0>0$ and $h_1>h_0$. The previous classification is in addition to the traditional division of time-periodic systems into slow ($\Omega_p<1$) and fast ($\Omega_p \geq 1$) modulation regimes \cite{Nassar201710}. Fig.~\ref{fig:Dispersion1_TimePeriodic} shows the dispersion patterns corresponding to the first category for various modulation regimes. Note that the effect of $h_0 \neq 0$ is a separation of the otherwise identical dynamic modes of the system as well as an up-shift of one of them by an amount $h_0/J \omega_0$ in the $\Omega-\kappa$ plot. Consequently, we only consider the zero-bias modulation amplitude category since the behavior of the remaining cases can then be readily extracted.

\begin{figure}[h!]
\includegraphics[width=0.5\textwidth]{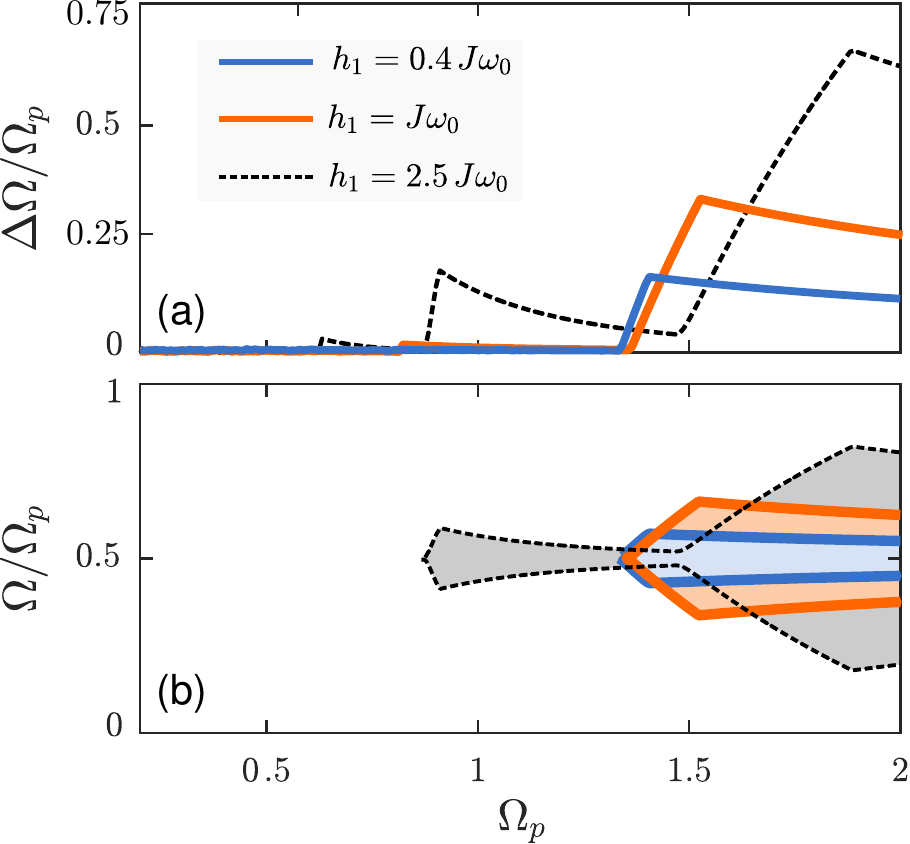}
\centering
\caption{(a) Band gap width $\Delta \Omega$ and (b) band gap frequency range in time-periodic GMMs with $h_0=0$}
\label{fig:Bandwidth}
\end{figure}

Note that the time-periodicity augments the dispersion profile with a periodicity along the frequency axis. That is, band diagrams at higher frequencies are repetitions of the frequency range $[0,\Omega_p]$. As a result, dispersion curves plotted within the ranges of $\kappa \in [-\pi,\pi]$ and $\Omega \in [0,\Omega_p]$ provide sufficient information about the propagation metrics of the time-periodic GMM. Furthermore, there does not exist any wavenumber band gaps (or complex $\lambda$ value with temporal instability \cite{cassedy1962temporal}) regardless of the speed or amplitude of the modulation. The latter indicates that a time-periodic GMM does not support unstable solutions with exponentially growing components as a result of complex eigenvalues. An observation which is in compliance with Eq.~(\ref{eq:Energy_TimePeriodic}) stating that the energy content of a time-periodic GMM is conserved over one modulation cycle. A detailed discussion of the stability of this class of GMMs will be separately addressed in a future effort. A parametric study on the effect of dimensionless pumping frequency $\Omega_p$ on the band gap width and range is depicted in Fig.~\ref{fig:Bandwidth} for the three cases shown in Fig.~\ref{fig:Dispersion1_TimePeriodic}. It is noted that the band gap width generally increases with the pumping frequency in every case, albeit not monotonically.

\subsection{Space-time-periodic Angular Momentum Variation: Breakage of Reciprocal Symmetry}

As outlined so far, a distinct advantage of GMM systems is the ability of an angular momentum (induced via a set of embedded spinning rotors) to influence elastic waves, onset band gaps and, more importantly, tune them in real-time by solely adjusting the rotation speed. In this section, we exploit such features to introduce a traveling-wave-like modulation of angular momentum with the intention of breaking the reciprocal symmetry of elastic waves in the GMM lattice. Consider the following space-time-periodic waveform for $h$:
\begin{equation} \label{eq:AngularMomentum_SpaceTimePeriodic}
    h^{(n)}(t)=h_0 + h_1 \cos(\omega_p t - \kappa_p n)
\end{equation}

\noindent for a GMM consisting of an infinite number of ``super cells", each of which containing $N$ unit cells with the prescribed time-varying angular momentum. Also, $n$ is the unit cell index and $\kappa_p=2 \pi/N$ is the spatial modulation frequency. It is worth noting that, unlike the time-periodic lattice, the angular momentum of adjacent unit cells here carries a $2 \pi /N$ phase difference. For convenience, we rewrite the governing motion equations for every $N$ gyric unit cells in the following LTP form
\begin{equation} \label{eq:Dynamics_SpaceTimePeriodic}
    \mathbf{J} \mathbf{\ddot{\Theta}} + \mathbf{G}(t) \mathbf{\dot{\Theta}} +
    \mathbf{K(\kappa)} {\mathbf{\Theta}}=0
\end{equation}
\noindent where $\mathbf{\Theta}_{2N \times 1}=[\theta_x^{(1)},...,\theta_x^{(N)}| \theta_y^{(1)},...,\theta_y^{(N)}]^{\text{T}}$ is the rotation vector, $\mathbf{J}=J \mathbf{I}$ is the inertia matrix,  and $\mathbf{I}$ is an identity matrix of size $2N$. In the absence of non-conservative constituents, $\mathbf{G}(t)$ takes the form
\begin{equation} \label{eq:GyroscopicMatrix_SpaceTimePeriodic}
\mathbf{G}(t)=
\left[
\begin{array}{c|c}
\mathbf{O} & \mathbf{H}(t) \\
\hline
-\mathbf{H}(t) & \mathbf{O}
\end{array} \right]_{2N \times 2N}
\end{equation}
\noindent from which it can be inferred that $\mathbf{G}(t)$ is a time-periodic skew-symmetric block matrix, i.e. $\mathbf{G}(t+\tau_p)=\mathbf{G}(t)=-\mathbf{G}^{\dag}(t)$ with $(\bullet)^{\dag}$ denoting the complex-conjugate-transpose operation. In Eq.~(\ref{eq:GyroscopicMatrix_SpaceTimePeriodic}), $\mathbf{H}(t)=\text{diag}\{h^{(1)}(t),h^{(2)}(t),...,h^{(N)}(t)\}$. Furthermore, the stiffness matrix $\mathbf{K}(\kappa)$ is explicitly
\begin{equation} \label{eq:Stiffness_SpaceTimePeriodic}
\mathbf{K}(\kappa)=
\left[
\begin{array}{c|c}
\mathbf{\Gamma(\kappa)} & \mathbf{O} \\
\hline
\mathbf{O} & \mathbf{\Gamma(\kappa)}
\end{array} \right]
\end{equation}
\noindent where the Hermitian matrix $\mathbf{\Gamma}(\kappa)$ is found to have the following general form  for $N \geq 3$:

\begin{align}
\mathbf{\Gamma}(\kappa)=k
\begin{bmatrix}
2 & -1 &  & -e^{i\kappa} \\
-1 & \ddots & \ddots & \\
& \ddots & \ddots &  -1 \\
-e^{-i\kappa} & & -1 & 2  \\
\end{bmatrix} 
\label{eq:Gamma_SpaceTimePeriodic}
\end{align}

\begin{figure*}
\includegraphics[width=\textwidth]{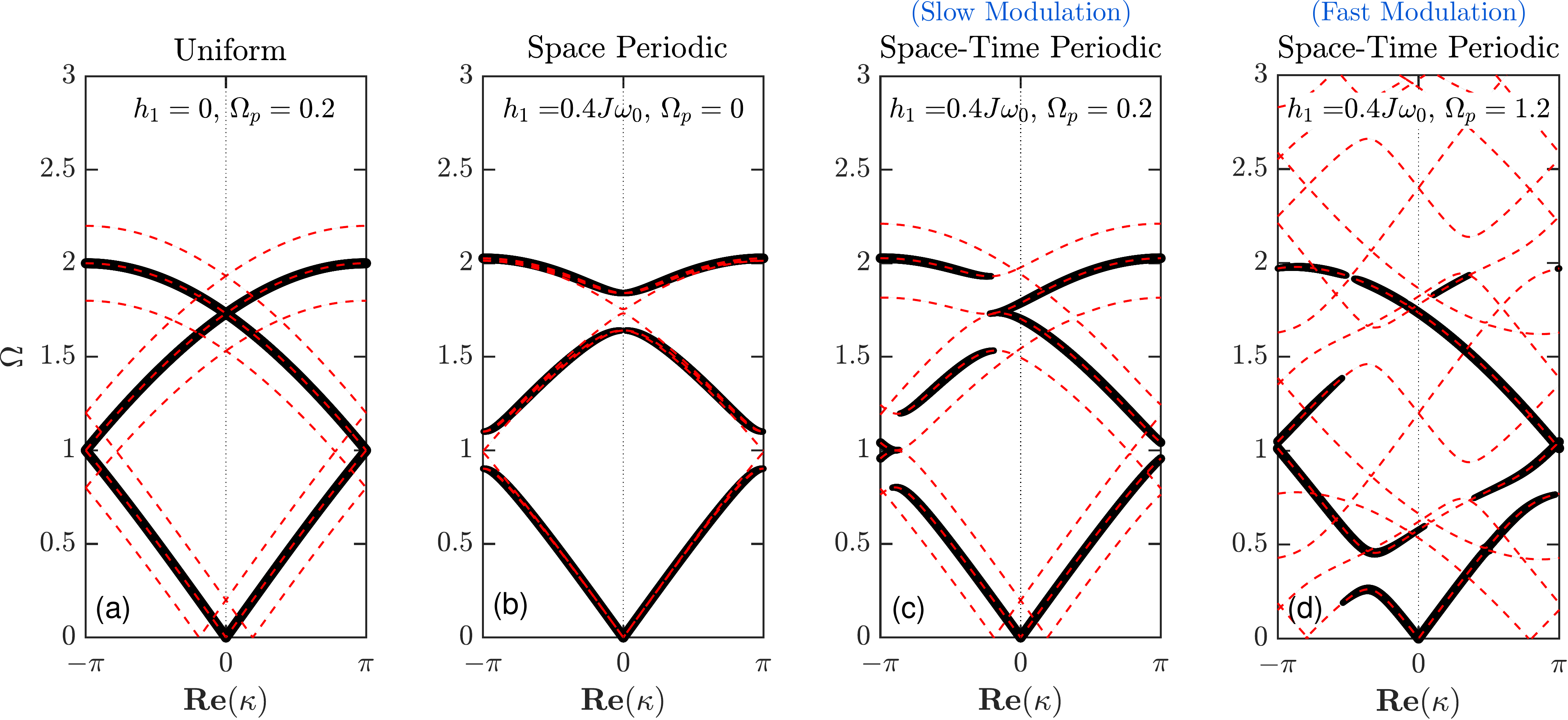}
\centering
\caption{Dispersion diagrams for uniform, space-periodic, and space-time-periodic GMMs with 3 unit cells per super cell (i.e. $N=3$) and a zero bias angular momentum value (i.e. $h_0=0$). Dashed lines denote all possible solutions of the theoretical dispersion analysis. Solid lines correspond to the fundamental modes obtained via an eigenvector weighting method}
\label{fig:DispersionN3_SpaceTimePeriodic}
\end{figure*}

In order to study propagation of linear elastic waves in the space-time-periodic GMM described by Eq.~(\ref{eq:Dynamics_SpaceTimePeriodic}), we build on the plane wave expansion method established for space-time modulated systems with $N \neq \infty$ \cite{Vila2017363,attarzadeh2018wave}. Correspondingly, we utilize a Bloch solution of the form $\mathbf{\Theta} (t)=\mathbf{\Phi}(t) e^{\lambda t}$ in which the amplitude vector $\mathbf{\Phi}(t)$ is $\tau_p$-periodic. As such, $\mathbf{\Phi}(t)$ and $\mathbf{G}(t)$ can be replaced with their Fourier series representations
\begin{equation} 
\mathbf{\Phi}(t)=\sum_{q=-\infty}^{\infty} \hat{\mathbf{\Phi}}_q e^{i q \omega_p t} \hspace{0.3cm} , \hspace{0.3cm}
\mathbf{G}(t)=\sum_{l=-\infty}^{\infty} \hat{\mathbf{G}}_l e^{i l \omega_p t}
\label{eq:FourierExpansions_SpaceTimePeriodic}
\end{equation}
\noindent where $\hat{\mathbf{\Phi}}_l$ and $\hat{\mathbf{G}}_q$ are the associated Fourier vector and matrix coefficients. By substituting in Eq.~(\ref{eq:Dynamics_SpaceTimePeriodic}) and exploiting the harmonic balance, we get
\begin{equation} \label{eq:DispersioRelation_SpaceTimePeriodic}
    (\mathbf{A} \Lambda^2 + \mathbf{B} \Lambda + \mathbf{C} 
    ) \mathbf{V}=0
\end{equation}
\noindent where $\mathbf{A}$, $\mathbf{B}$ and $\mathbf{C}$ are block matrices with entries that are explicit functions of $N$, $\Omega_p$, $h_0/J\omega_0$ and $h_1/J\omega_0$ (see supplementary material for details \cite{supp}). In addition, the vector $\mathbf{V}$ is obtained by stacking all the $\hat{\mathbf{\Phi}}_{q}$ from $q=-\infty$ to $\infty$ as follows
\begin{equation}
    \mathbf{V}^{\text{T}}=[\hat{\mathbf{\Phi}}_{-\infty}^{\text{T}},...,\hat{\mathbf{\Phi}}_{-1}^{\text{T}},\hat{\mathbf{\Phi}}_{0}^{\text{T}},\hat{\mathbf{\Phi}}_{1}^{\text{T}},...,\hat{\mathbf{\Phi}}_{\infty}^{\text{T}}]
\end{equation}

Solving the eigenvalue problem in Eq.~(\ref{eq:DispersioRelation_SpaceTimePeriodic}) yields dispersion diagrams of the space-time-periodic GMM. For any given $\kappa$, only $2(2N)$ solutions are fundamental modes that can be identified using the magnitude of the associated eigenvector components \cite{Vila2017363} (see supplementary material for more on dispersion band reduction \cite{supp}).

Fig.~\ref{fig:DispersionN3_SpaceTimePeriodic} shows the dispersion plots for a uniform, space-periodic, and space-time-periodic (with two different modulation speeds) for a GMM with 3 gyric unit cells per super cell (i.e. $N=3$) and $h_0=0$. The dispersion behavior in the first two cases is fully reciprocal, with band gaps emerging in Fig.~\ref{fig:DispersionN3_SpaceTimePeriodic}b due to the spatial periodicity. By imposing a temporal modulation with a frequency $\Omega_p=0.2$, the GMM's transmission symmetry is broken and the dispersion becomes non-reciprocal (Fig.~\ref{fig:DispersionN3_SpaceTimePeriodic}c). Note that band gaps corresponding to left-propagating waves up-shift a/o down-shift by an amount $\Omega_p$; a behavior which is in agreement with the observed robustness \cite{chaunsali2016stress} and quantization \cite{nassar2018quantization} of non-reciprocal band gaps in space-time modulated systems. As indicated earlier, the GMM uniquely maintains stability at fast modulation speeds, in contrast to stiffness modulation in elastic metamaterials \cite{attarzadeh2018wave}. An example of a fast modulation is $\Omega_p=1.2$ in Fig.~\ref{fig:DispersionN3_SpaceTimePeriodic}d. The result is a much greater degree of distortion and asymmetry in the dispersion curves, albeit without wavenumber band gaps or complex frequencies. Finally, it is noted that 10 harmonic modes are included in Fig.~\ref{fig:DispersionN3_SpaceTimePeriodic}d, whereas one mode sufficiently predicts the dispersion behavior with a reasonable accuracy in the low-speed modulation case.

\section{Numerical Verification}
\begin{figure*}
\includegraphics[width=\textwidth]{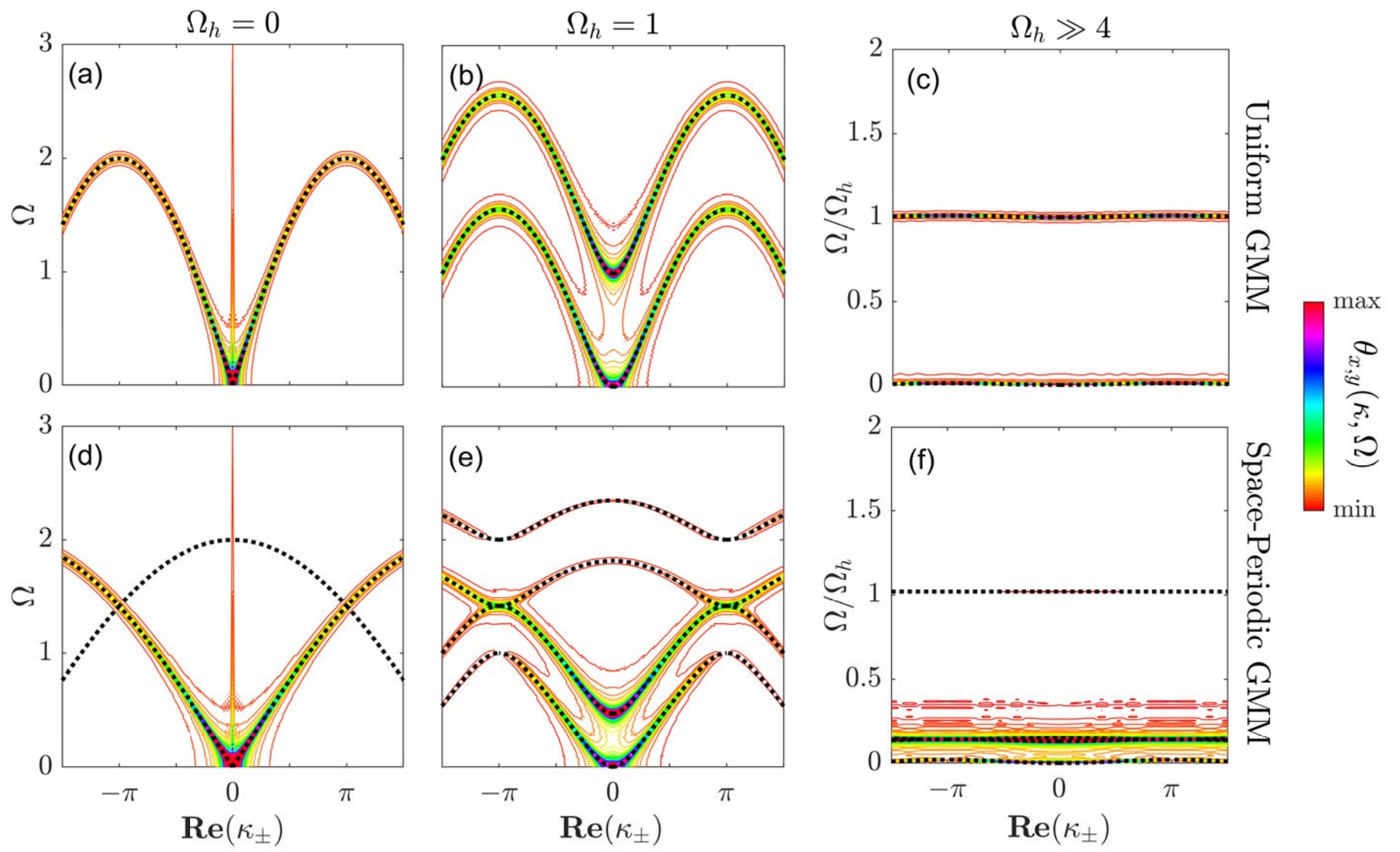}
\centering
\caption{Numerically reconstructed dispersion contours for a GMM with (a-c) uniform and (d-f) space-periodic angular momentum distribution. Dotted lines denote the theoretically predicted dispersion bands for comparison}
\label{fig:UniformAndSpacePeriodic_Numerical}
\end{figure*}

To verify the dispersion predictions detailed in Section~\ref{Wave Dispersion in gyric-metamaterial}, the transient response of a finite 1D GMM lattice with 300 cells is numerically computed via full wave simulation and is used to evaluate the GMM's actual dispersion contours. The GMMs in the simulations are subjected to a broadband force imparted at the center of the lattice such that both forward and backward traveling waves can be captured simultaneously. The excitation is a Gaussian wave packet whose frequency content spans the frequency range of interest. We solve for the time-dependent displacement field of the lattice using MATLAB's ``ode45" function with 1e-6 relative tolerance and an output time step of 5 ms. A Hamming window is also incorporated to compensate for the truncation of signal to the first 15 seconds---before it reaches spatial boundaries. A two-dimensional Fourier transformation (2DFT) is then applied in both space and time to numerically reconstruct the dispersion diagrams of each individual case. The 2DFT is given by
\begin{equation}
    \theta(\kappa,\Omega)=\sum_{n=1}^{300} e^{i\kappa \frac{n}{N}} \int_{0}^{t_{\text{end}}} \theta^{(n)}(t) \,e^{-i\omega_0 \Omega t} \,dt
\end{equation}
where $t_{\text{end}}$ is the total simulation time. In here, a grid of $200\times200$ is used to represent the transformed displacement amplitude in the $\kappa-\Omega$ space. Figs.~\ref{fig:UniformAndSpacePeriodic_Numerical}a-c graphically display the recovered dispersion behavior corresponding to a finite GMM lattice with time-invariant uniform distribution of angular momentum. The numerical contours are in strong agreement with the theoretical dispersion predictions, plotted via dashed lines for comparison. Note that as $\Omega_h$ increases from zero, each dynamic modes of the system splits into two separate wave-modes, as shown in Fig.~\ref{fig:UniformAndSpacePeriodic_Numerical}b for $\Omega_h=1$. By further increasing the speed of the spinning rotors, they converge to two flat modes at $\Omega=0$ and $\Omega=\Omega_h$ effectively suppressing almost the entire frequency spectrum. 

The second set of simulations, shown in Figs.~\ref{fig:UniformAndSpacePeriodic_Numerical}d-f, correspond to the spatially periodic gyric lattice which comprises a zero angular momentum for every other unit cell. In this case, the two initially repeated wave-modes of the non-gyric lattice (Fig.~\ref{fig:UniformAndSpacePeriodic_Numerical}d) turn into four distinct wave-modes (Fig.~\ref{fig:UniformAndSpacePeriodic_Numerical}e) as $\Omega_h$ grows from zero to one. By further increasing $\Omega_h$, the two middle wave-modes coincide to form a single standing-wave (flat) mode at $\Omega_h=\sqrt{2}$ while the two other bands become nearly flat at $\Omega=0$ and $\Omega_h$, with the number of unique dynamic modes reduced to three (Fig.~\ref{fig:UniformAndSpacePeriodic_Numerical}f).  Once again, the GMM becomes an almost all-attenuation structure suppressing the bulk of the frequency spectrum. Interestingly in this case, the wave mode emerging at $\Omega=\sqrt{2}$ appears to be dominant (since it's a coalescence of two modes as evident by the contour amplitudes).

\begin{figure*}
\includegraphics[width=\textwidth]{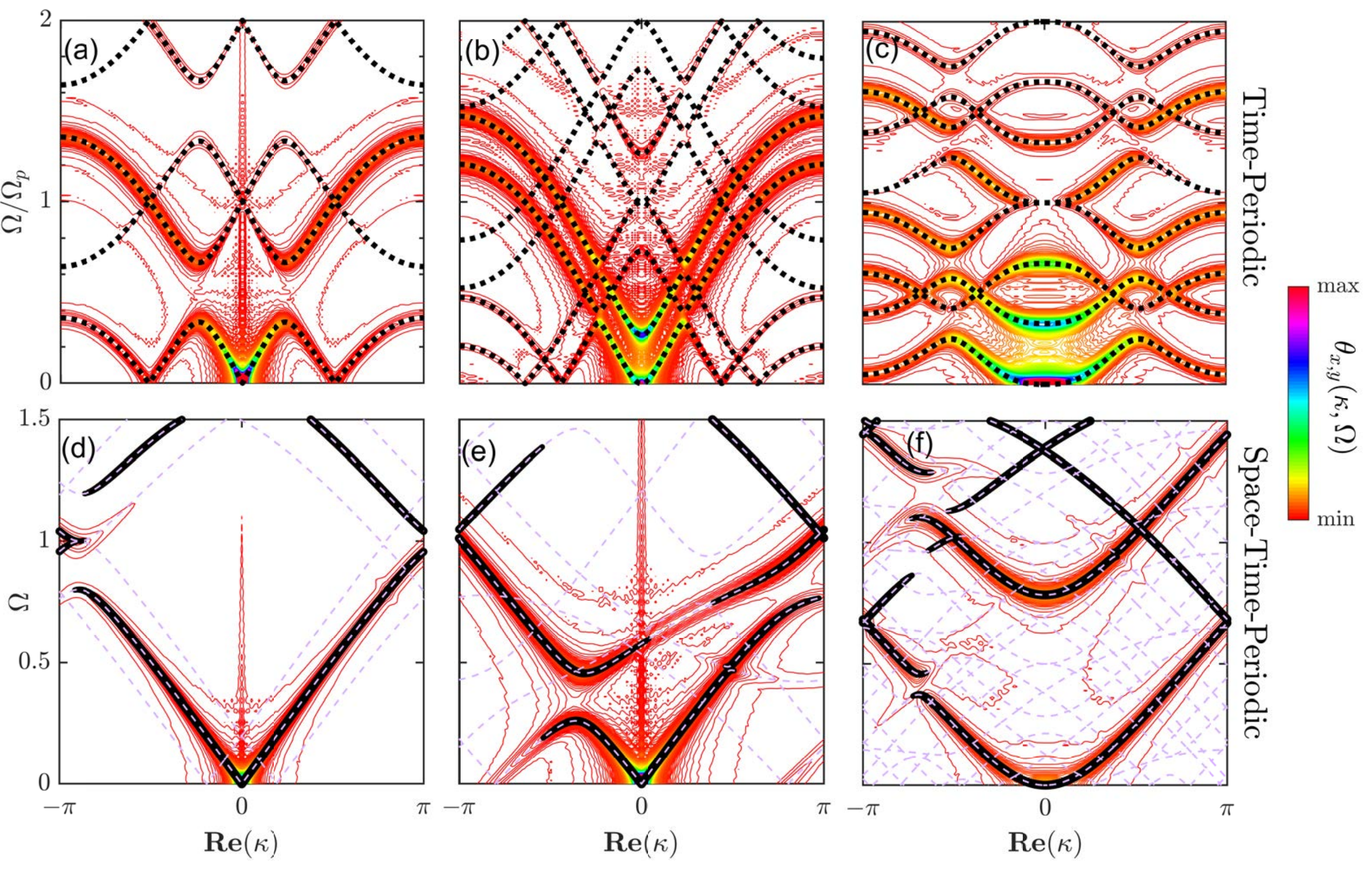}
\centering
\caption{Numerically reconstructed dispersion contours for a time-periodic GMM (a-c) and a space-time-periodic GMM (d-f) with the following angular momentum modulations: (a) $h_0=0$, $h_1=J \omega_0$, and $\Omega_p=1.5$, (b) $h_0=0.4J\omega_0$, $h_1=0.5 h_0$, , and $\Omega_p=1.5$, (c) $h_0=2.5 J\omega_0$, $h_1=1.5 h_0$, and $\Omega_p=1.5$, (d) $h_0=0$, $h_1=0.4 J\omega_0$, and $\Omega_p =0.2$, (e) $h_0=0$, $h_1=0.4 J\omega_0$, and $\Omega_p =1.2$, and (f) $h_0=0.8 J \omega_0$, $h_1=\frac{3}{8} h_0$, and $\Omega_p =0.5$. Theoretical dispersion curves are shown for comparison}
\label{fig:SpaceTimePeriodic_Numerical}
\end{figure*}

In keeping with the previous cases, numerical simulations are used to validate the dispersion predictions for the time-periodic GMM. Fig.~\ref{fig:SpaceTimePeriodic_Numerical} shows the behavior of the time-periodic GMM for select modulation cases at $\Omega_p=1.5$. Fig.~\ref{fig:SpaceTimePeriodic_Numerical}a corresponds to $h_0=0$ and $h_1=J \omega_0$. The repetition of the first extended Brillouin zone is clearly noticeable. As the theory suggests, one band gap appears in each extended Brillouin zone that is repeated periodically at higher frequencies. Figs.~\ref{fig:SpaceTimePeriodic_Numerical}b and c are obtained for $h_0=0.4J\omega_0, \hspace{0.1cm} h_1=0.5 h_0$, and $h_0=1.5J\omega_0, \hspace{0.1cm} h_1=2.5 h_0$, respectively. A near perfect agreement between the theoretical predictions and the simulations can also be observed throughout. Finally, Fig.~\ref{fig:SpaceTimePeriodic_Numerical}d-f shows the simulation results for the space-time-periodic GMMs. Dashed lines correspond to all the dispersion bands obtained from the theoretical analysis, while the solid black lines represent the modes with greatest weights. In all the three modulation cases shown, reciprocity is broken and waves behave differently on both sides of $\kappa=0$.

\vspace{-0.3cm}

\section{Conclusions}
This work has showcased gyric metamaterials (GMMs) as an alternative class of metamaterials which can effectively exhibit wave manipulation capabilities culminating in band gaps as well as non-reciprocal transmission without necessarily requiring a mass or stiffness modulation. By tuning the rotational speed or direction of a set of embedded spinning rotors, a wide range of spatiotemporal distributions of the angular momentum of the lattice's unit cells can be achieved. Such variations have been shown to onset band gaps in the frequency spectrum and/or induce one-way directivity as needed.  A combination of theoretical dispersion analyses as well as numerical time-domain simulations have been utilized to show the ability of the angular momentum modulation to achieve some of the fundamental and most recent developments in non-reciprocal systems and metamaterial-based wave guides. Unlike inertia and stiffness, angular momentum is not an inherent property of the material and can, therefore, be handily tuned by solely controlling the speed of rotors. As such, a gyric lattice can near-instantaneously switch between a set of different desirable functionalities by design. Such tunability becomes even more significant in the context of non-reciprocal wave phenomena and space-time-periodic systems, where a certain degree of online control over the system's mechanical properties is critical. Further, given their dependence on gyroscopic effects, in a frictionless setup, GMMs can be classified as conservative systems which do not dissipate mechanical energy into heat or other disorganized energy forms. Finally, the analysis undertaken here illustrated that non-reciprocal GMMs remain reliably stable at high pumping frequencies. Future directions of this work include an experimental realization of this class of time-periodic gyric lattices as well as a rigorous Lyapunov-based investigation of the different stability metrics of such GMMs.

\vspace{-0.3cm}

\section{Acknowledgement}

M. A. and M. N. acknowledge the support of this work from the Vibration Institute (VI) through the 2018 VI Academic Grant Program.

\bibliography{References}

\end{document}